\documentclass[journal]{IEEEtran}

\usepackage{cite}      

\usepackage{graphicx}  
\begin{document}
%

\title{The Effects of Preheating of a Fine Tungsten Wire\\ and the Polarity of a High-Voltage Electrode\\ on the Energy Characteristics of an Electrically Exploded Wire in Vacuum}
%
%
\author{A.\,G.~Rousskikh,
        R.\,B.~Baksht,
        S.\,A.~Chaikovsky,
        A.\,V.~Fedunin,
        K.\,V.~Khishchenko,
        A.\,Yu.~Labetsky,
        P.\,R.~Levashov,
        A.\,V.~Shishlov,
        and~S.\,I.~Tkachenko
\thanks{A.\,G.~Rousskikh, R.\,B.~Baksht, S.\,A.~Chaikovsky, A.\,V.~Fedunin, A.\,Yu.~Labetsky, and A.\,V.~Shishlov 
are with the Institute of High Current Electronics, 
Siberian Branch of Russian Academy of Sciences,
Akademicheskiy Prospekt 2/3, Tomsk 634055, Russia (e-mail: russ@ovpe2.hcei.tsc.ru).}%
\thanks{K.\,V.~Khishchenko, P.\,R.~Levashov, and S.\,I.~Tkachenko are with the Institute for High Energy Densities, 
Joint Institute for High Temperatures, 
Russian Academy of Sciences,
Izhorskaya 13/19, Moscow 125412, Russia (e-mail: svt@ihed.ras.ru).}
}
%
%
%
\markboth{Journal of \LaTeX\ Class Files,~Vol.~1, No.~11,~August~2006}{Shell \MakeLowercase{\textit{et al.}}: Bare Demo of IEEEtran.cls for Journals}
%



\maketitle

\begin{abstract}
Results obtained from experimental and numerical studies of tungsten wires electrical explosion in vacuum are presented. 
The experiments were performed both with and without preheating of the wires, 
using positive or negative polarity of a high-voltage electrode. Preheating is shown to increase energy deposition in the wire core due to a longer resistive heating stage. 
The effect was observed both in single wire and wire array experiments. The evolution of the phase state of the wire material during explosion was examined by means 
of one-dimensional numerical simulation using a semiempirical wide-range equation of state describing the properties of tungsten with allowance made for melting and vaporization.\end{abstract}


%
\IEEEpeerreviewmaketitle

\section{Introduction}

\PARstart{E}{lectrical} explosion of metal wire arrays is the basis for production of high-power pulsed soft x-ray radiation in wire array implosion experiments
\cite{Spielman:PP:1998,Lebedev:PP:2001,Volkov:PPR:2004}. In experiments of this kind, the x-ray power depends, among other factors, 
on the time of the wire material transition from a solid to a gas-plasma state. In actual situations, a shunting breakdown of the electrode gap impedes 
complete wire vaporization (see, e.g., papers \cite{Alexandrov:PPR:2003,Rousskikh:PP:2004} and references therein). After the breakdown, a low-density corona is generated around the wires. 
Part of the current is carried through the corona plasma accelerated toward the axis of the system as jets formed from each wire \cite{Lebedev:PP:2001,Alexandrov:IEEETPS:2002}. 
In this way, a rarefied hot plasma retarding wire array implosion is formed long before the bulk of the wire array material arrives at the axis of the system. 
Moreover, part of the wire array mass is never involved in implosion and remains at the periphery, with part of the current from the generator also flowing 
through the residual plasma \cite{Alexandrov:HPB2004}. From this it follows that the development of the shunting breakdown must be retarded to improve the wire array implosion dynamics.

\section{Experiment and Simulation}

We have investigated fine wire explosion, using positive or negative polarity of the high-voltage electrode of a current generator. 
Emphasis was placed on elucidation of the causes of the shunting discharge along an electrically exploded fine wire.

The experiments were carried out on the current generator described previously \cite{Rousskikh:Tomsk:2004}. 
Capacitance and inductance of the generator are 70~nF and 730~nH correspondingly. In the experiments under consideration, 
tungsten wires 30.5~$\mu$m in diameter and $20\pm 0.5$~mm in length were exploded. 

Use was made of the following electrophysical diagnostics: a resistive divider, 
a shunt placed on the ground electrode side, and a bare x-ray vacuum diode with an aluminum photocathode. The bandwidth of the resistive divider and the current shunt is not worse than 300~MHz.
A framing camera was used for optical diagnostics. The camera was gated by a voltage pulse from the high voltage electrode. 
The exposure time was determined by the interval, during which the voltage across the anode--cathode gap was 7~kV and higher. The gated pulse from the high voltage electrode to the framing camera was delayed on the time $\Delta t$, which depends on the length of a cable connecting the high voltage electrode and the framing camera. This enabled us to record the onset of the electrode gap breakdown and to cut off bright light flashes inherent in the arc phase of the shunting discharge. The system has an undoubted benefit of automatic synchronization of the framing camera operation with the wire explosion. 

The experiments were carried out in four wire explosion modes, which were determined by the polarity of the high-voltage electrode and the use of wire preheating: 
mode A --- the high-voltage electrode served as a cathode, $U_0=10$, 20 or 30~kV, no wire preheating; 
mode B --- the high-voltage electrode served as an anode, $U_0=10$, 20 or 35~kV, no wire preheating; 
mode C --- the high-voltage electrode served as a cathode, $U_0=10$ or 30~kV, a wire was preheated; 
mode D --- the high-voltage electrode served as an anode, $U_0=10$ or 35~kV, a wire was preheated. 
Here, $U_0$ is the capacitor charging voltage. The wire was preheated to $T_0=1.4$--1.6~kK maintained for one hour, using a direct current of 150~mA. 
Because heating was not switched off during the explosion, the heater supply circuit was protected by a 1.1~mH inductance.

\begin{figure}[h]
\centering
  \includegraphics[clip=true,width=82mm]{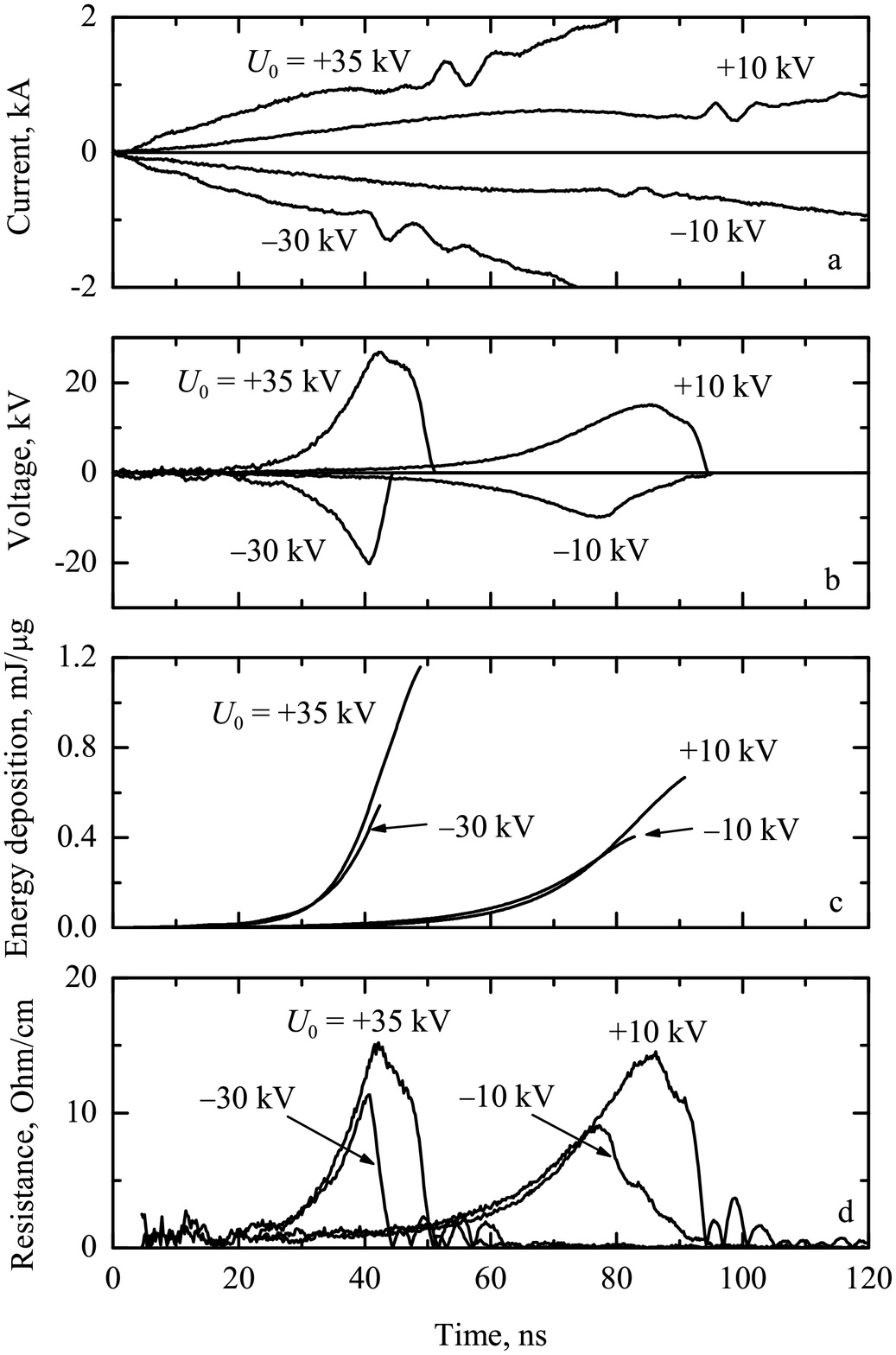}
  \caption{
Temporal variations of the current (a) and voltage (b) applied to an initially cold tungsten wire 30.5~$\mu$m in diameter, 
energy deposition in the wire (c), and resistance of the cathode--anode gap with a wire load (d) for different polarities of the high-voltage electrode.
}
  \label{osc1}
\end{figure}

\begin{figure}[h]
\centering
  \includegraphics[clip=true,width=82mm]{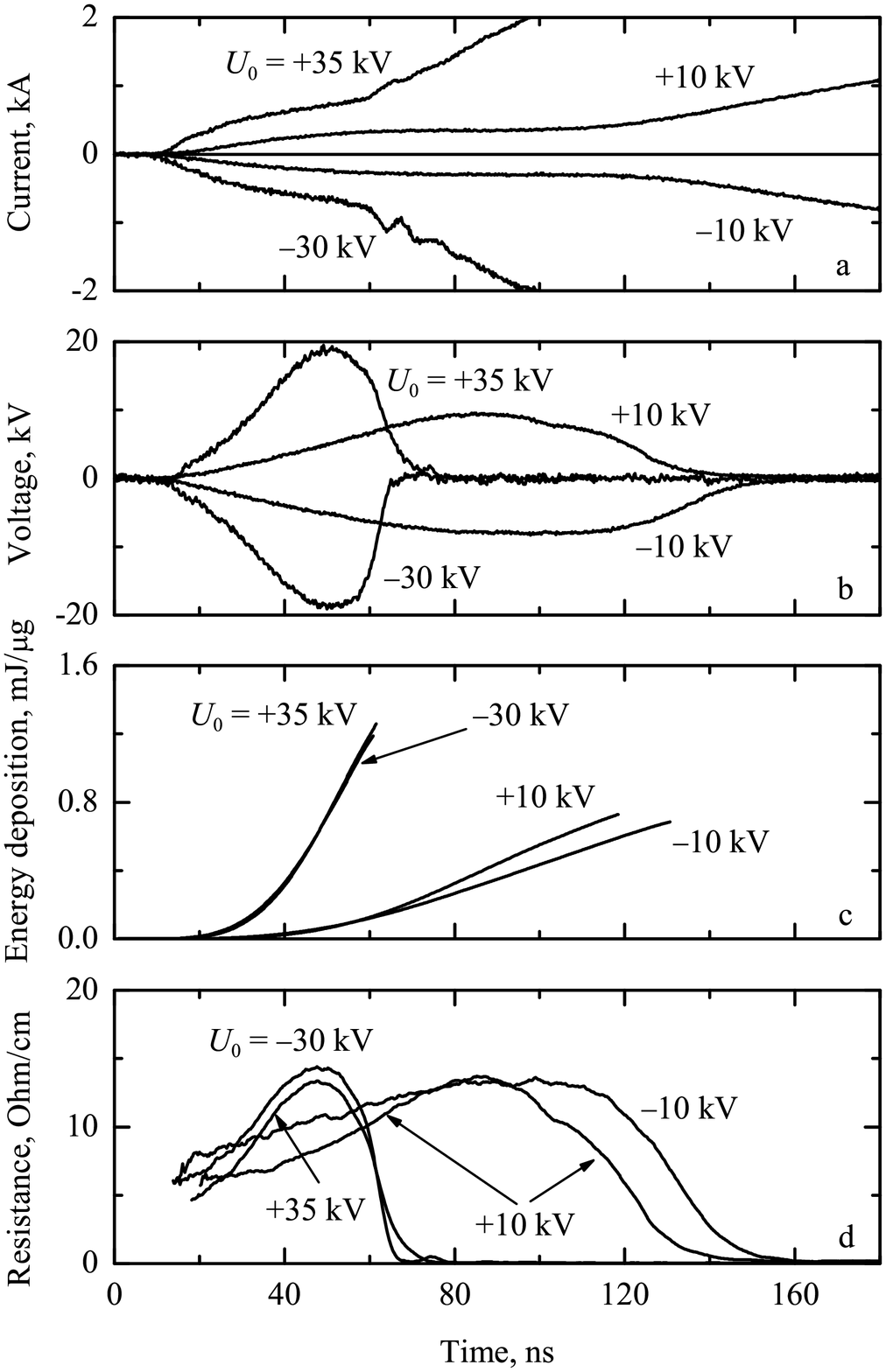}
  \caption{
Temporal variations of the current (a) and voltage (b) applied to a preheated tungsten wire 30.5~$\mu$m in diameter, 
energy deposition in the wire (c), and resistance of the cathode--anode gap with a wire load (d) for different polarities of the high-voltage electrode.
}
  \label{osc2}
\end{figure}

\begin{figure}
\centering
  \includegraphics[width=82mm]{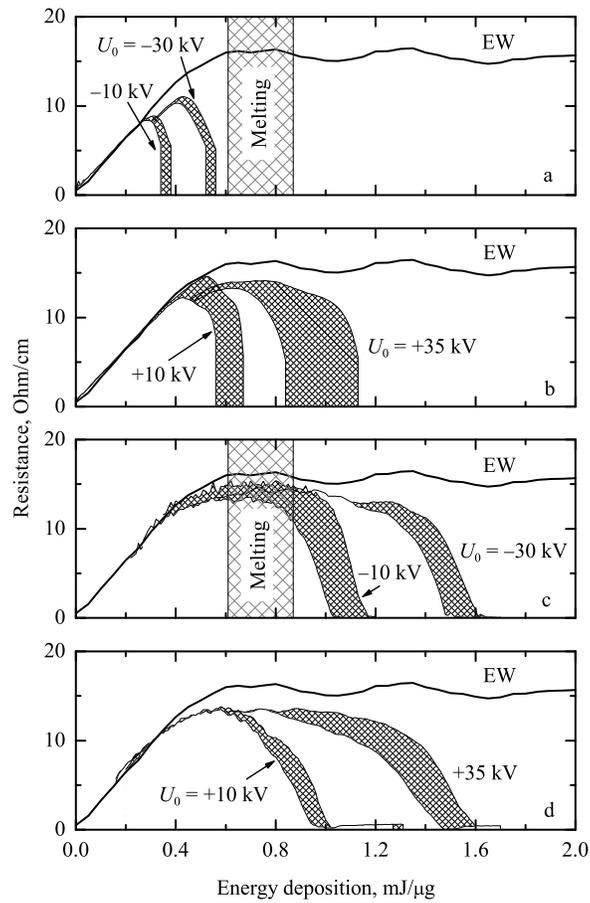}
  \caption{
Resistance of the cathode--anode gap with a wire load versus energy deposition: 
(a) explosion mode A, the high-voltage electrode is a cathode, no preheating; 
(b) explosion mode B, the high-voltage electrode is an anode, no preheating; 
(c) explosion mode C, the high-voltage electrode is a cathode, a wire is preheated; 
(d) explosion mode D, the high-voltage electrode is an anode, a wire is preheated.
The shaded regions correspond to statistical spread in the data. 
EW denotes experimental data for explosion of tungsten wire with 30.5~$\mu$m diameter in water \cite{Rousskikh:Tomsk:2004}.
Melting boundaries are in accordance with results of measurements \cite{Hixson:IJT:1990}.
}
  \label{RW}
\end{figure}

Figures~\ref{osc1} and \ref{osc2} show current, voltage, energy deposition, and resistance waveforms obtained 
in the experiments where a maximum peak voltage across the wire was recorded
in different electrical explosion modes. 
Calculation of the deposited energy (see Fig.~\ref{osc1}c and \ref{osc2}c) was stopped when 
the resistive voltage 
of the cathode--anode gap drops two times from its maximum value. 
As it was already demonstrated experientially in \cite{Sarkisov:PRE:2002,Duselis:PP:2004}, 
the polarity of the high-voltage electrode has a profound effect on energy deposition in the wire. It is evident from the oscillograms presented in Fig.~\ref{osc1} 
that it takes much less time to produce a breakdown along a wire in the case of negative polarity of the high-voltage electrode than of positive polarity. 
Figure~\ref{osc2} shows similar oscillograms taken during a preheated wire explosion. It is obvious that the polarity of the high-voltage electrode has only 
a marginal effect in those experiments. This fact testifies that surface impurities and desorbed gases play a key role in the formation of the breakdown along the wire 
responsible for a lower energy deposited in the wire core. 
To visualize spread in the experimental data, we demonstrate generalized results as a maximum and a minimum 
resistance of a tungsten wire 30.5~$\mu$m in diameter in relation to energy deposition (Fig.~\ref{RW}). 
A minimum spread in the experimental data is observed in the case of negative polarity of the high-voltage electrode without wire preheating.

Analysis of the results obtained from wire explosion experiments without preheating for different polarities of the high-voltage electrode (modes A and B) 
shows that positive polarity provides higher energy deposition than negative polarity. A comparison of the results for modes C and D suggests that there is a minor difference 
in energy deposition.

\begin{figure}
  \begin{center}
    \includegraphics[width=40mm,trim=0 0 0 35,clip=true]{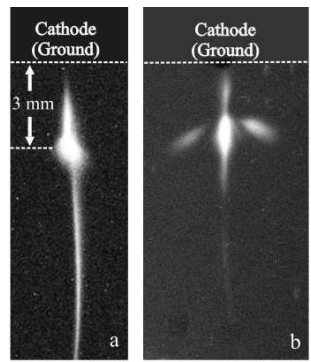}
  \end{center}
  \caption{Framing camera images of the near-cathode region of the discharge gap for a tungsten wire explosion 
with positive polarity of the high-voltage electrode (mode B, $U_0=20$~kV).
The delay time is 3.5~ns (a) and 25~ns (b). The exposure is 14~ns (a) and 23~ns (b).}
  \label{cathodeplus}
\end{figure}

\begin{figure}
  \begin{center}
    \includegraphics[width=78mm,trim=0 0 0 0,clip=true]{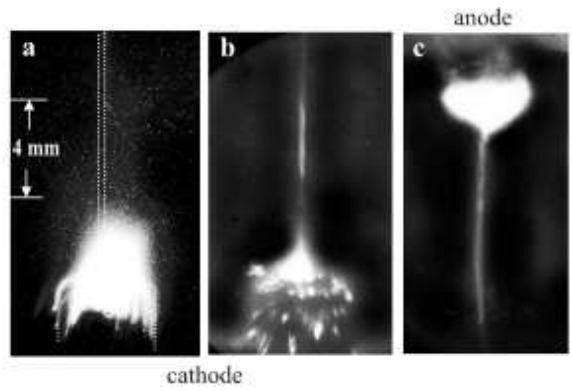}
  \end{center}
  \caption{Framing camera images of the near-cathode region (a), (b) and the near-anode region (c) of the discharge gap 
for a tungsten wire explosion with negative polarity of the high-voltage electrode (mode A, $U_0=20$~kV). 
The delay time is 3.5~ns (a) and 57~ns (b), (c). The exposure is 8~ns (a), (b) and 14~ns (c).}
  \label{cathodeminus}
\end{figure}

Judging by the framing camera images, we may conclude that the duration of the pre-breakdown phase depends on special features of plasma formation in the near-cathode region. 
For positive polarity of the high-voltage electrode (cathode serves as the ground electrode), an emission center strongly connected with the development of the breakdown emerges on the 
wire surface $\sim 3$~mm far from the cathode (Fig.~\ref{cathodeplus}a). Light emission from this center becomes much brighter than that from the wire 25~ns after the breakdown 
(Fig.~\ref{cathodeplus}b), 
light emission from the surface of the electrodes being not observed. 
For negative polarity of the high-voltage electrode, breakdown is strongly connected with an emergence of emission centers 
immediately on the cathode surface (Fig.~\ref{cathodeminus}a, b) and with an uniform high-intensity light emission region on the wire
surface $\sim 4$--6~mm far from the cathode. 
The number of emission centers on the cathode 
may vary from shot to shot. Early in the breakdown process, the emission centers formed on the end face of the cathode occupy the entire surface of the high-voltage electrode 
in the course of time. Uniform high-intensity light emission on the electrode surface is observed from the anode side (ground electrode), as shown in Fig.~\ref{cathodeminus}c.

We have looked at the characteristics of individual wires in an electrically exploded 12~mm wire array comprising 9 tungsten wires 30.5~$\mu$m in diameter. 
The wire array was exploded for negative polarity of the high-voltage electrode with $U_0 = 20$~kV. Both initially cold (room-temperature) and preheated wire arrays were used. 
The wires were preheated to a temperature of $T_0=1.8$~kK maintained for one hour.

\begin{figure}
\centering
    \includegraphics[width=82mm]{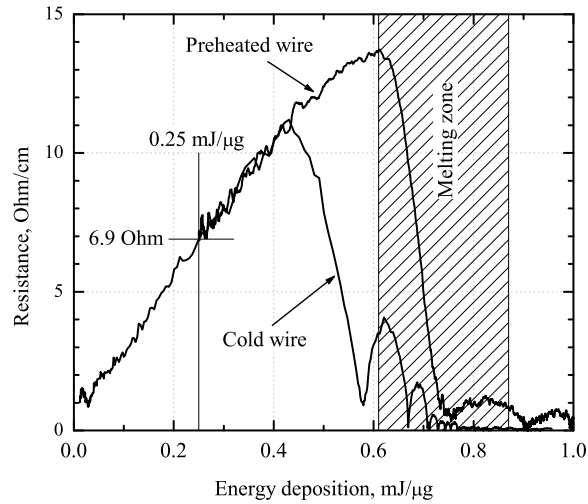}
  \caption{Resistance per unit length of an individual tungsten wire with 30.5~$\mu$m diameter in an array of 12 wires.}
  \label{indwire}
\end{figure}

The initial resistance of a preheated wire array is much higher than that of a cold one. Since the current generator is started up at varying initial load resistance and is operated 
in different modes, a comparative analysis of the oscillograms of current and voltage waveforms is difficult to perform. It is much more convenient to consider the resistance 
of individual wires in an array as a function of energy deposition (RW diagram). Figure~\ref{indwire} illustrates variations of the resistance per unit length of an individual wire 
in an array with energy deposition in the wire. In a preheated wire array, the initial resistance of an individual wire was 6.9~Ohm, which corresponded to an initial wire temperature of 1.8~kK 
and deposited energy of 0.25~mJ/$\mu$g. Hence, the plot for a preheated wire given in Fig.~\ref{indwire} begins with these data point. 
The results cited lead us to conclude that for negative polarity the energy deposited in a preheated wire array is somewhat higher than in a cold wire array. 
This conclusion is supported by the experimental fact that the cold wire array was either unaffected by exposure to generator current or only one wire of the array was burnt out. 
In the preheated wire array, current flow through the wire load caused destruction of all wires. The maximum resistance was 1.3--1.4 times larger for the preheated wire array 
than for the originally cold one. This is also evidence of an increase in the energy deposited both in individual wires and in the whole wire array. It should be noted 
that when the generator current is passed through a cold wire array, the onset of the shunting breakdown (see Fig.~\ref{indwire}) corresponds to the lower bound 
of statistical spread of the curves in the RW diagram for an individual wire (see Fig.~\ref{RW}a).

\begin{figure}
\centering
    \includegraphics[width=82mm]{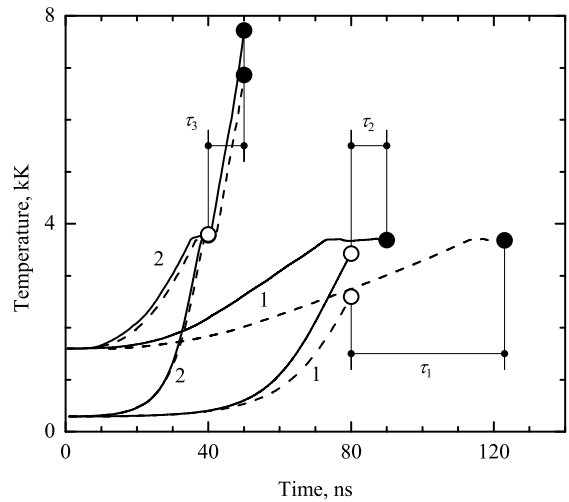}
  \caption{
Tungsten wire temperature versus heating time: calculated explosion modes with positive (solid lines) and negative polarity (dashed lines), 
the onset of the electric breakdown for preheated (full circles) and cold wires (open circles), $U_0 = 10$~kV (1) and 30 or 35~kV (2).
}
  \label{simulation}
\end{figure}

We have calculated the resistive stage of the electrical explosion of a tungsten wire 
using experimental current waveform and breakdown time 
within a one-dimensional magnetohydrodynamic model following the computational 
procedure described in \cite{Tkachenko:TVT:2001}. The thermodynamic properties of tungsten were determined using a wide-range multi-phase equation of state 
\cite{Khishchenko:Elbrus2005:EOS} in tabulated form \cite{Levashov:Elbrus2004:Tabular}. The electrical conductivity was approximated by a semiempirical relation \cite{Knoepfel:1970}. 
The wire preheating level was assumed to be the same in explosion modes C and D with the wire temperature being $T_0=1.6$~kK. 
Figure~\ref{simulation} shows results of numerical calculations for the initial explosion phase in different wire explosion modes.

According to experimental data, we can say that the time of shunting breakdown is delayed by $\tau_1\sim 40$~ns in the case of negative polarity of the high-voltage electrode 
and a preheated wire with $U_0 = 10$~kV as compared to the results obtained from the cold wire experiments. According to our numerical calculations the preheated wire 
temperature immediately prior to the breakdown was estimated to be higher by $\Delta T_a \sim 1$~kK than that of the cold wire. For positive polarity, both the delay in the breakdown 
and further increase in temperature were smaller than for negative polarity: $\tau_2 \sim 10$~ns and $\Delta T_a \sim 0.4$~kK 
(in this case, part of the energy deposited in a wire was spent for melting). In explosion modes with faster energy deposition ($U_0 = 30$~kV or 35~kV) 
the delay depends only weakly on polarity, $\tau_3 \sim 10$~ns. In preheated wire explosion modes, the wires are melted completely and, what is more, are heated further up to 
$T \sim 8$~kK thereafter with $\Delta T_a \sim 4$~kK. Experimental results reported in \cite{Ivanenkov:TP:1995,Pikuz:PRL:1999} also point to the fact that preheating enables a higher energy 
to be deposited in a tungsten wire in the resistive heating phase.

Thus, we can assume that gas desorption from the wire surface during explosion may be substantially reduced by preheating a wire array. As a consequence, 
the heating time during explosion is increased, resulting in a higher energy deposited in the wire core. This may have a profound effect on the x-ray radiation spectrum and intensity 
in wire array implosion. To verify this assumption, we have performed proof-of-principle experiments on preheated and cold wire array implosion using the GIT-12 generator. 
The load was an array of 12 tungsten wires of 11 $\mu$m in diameter arranged in a circle of 21~mm in diameter. The wire array length and mass were 15~mm and 220~$\mu$g/cm correspondingly. 
The wires were subjected to preheating for one hour until the current generator was triggered. The resistance of a preheated wire array  was 4.5~Ohm, which corresponded to a temperature of 1.88~kK.

The following diagnostics were used in the experiments: x-ray vacuum diode (XRD1) with an aluminum photocathode filtered by a 4~$\mu$m Kimfoil plus 0.4~$\mu$m aluminum filter, another x-ray vacuum diode (XRD2) with an aluminum photocathode behind a 3~$\mu$m Mylar filter, and a streak camera with a writing speed of 125~ns/cm. The slit of the camera was normal to the wire array axis.

\begin{figure}
\centering
    \includegraphics[width=85mm]{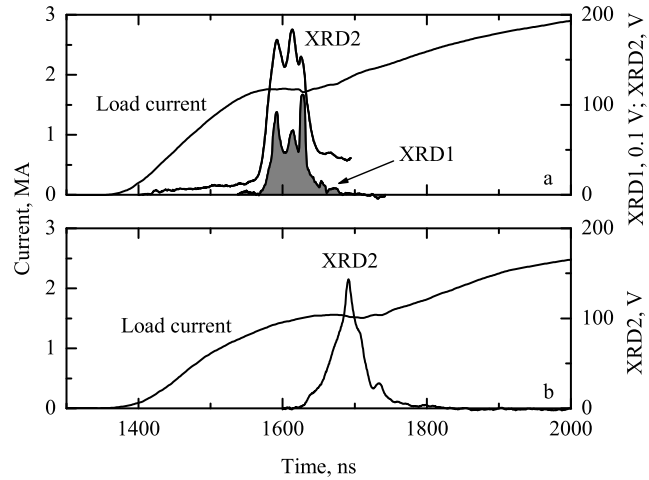}
  \caption{
Temporal variations of the generator current and XRD1 and XRD2 signals: an array of 12 tungsten wires 11~$\mu$m in diameter with preheating (a) and without preheating (b).
}
  \label{temporal}
\end{figure}

Figure~\ref{temporal} shows oscillograms of the generator current and XRD1 and XRD2 output signals. For an initially cold electrically exploded wire array, 
no response from XRD1 was detected. This fact suggests that radiation from a cold wire array lies in an energy range below 200~eV. In preheated wire array experiments, 
the radiation spectrum was found to shift to a shorter wavelength region as evidenced by an appreciable signal from XRD1 (see Fig.~\ref{temporal}a).

\begin{figure}
  \begin{center}
    \includegraphics[width=88mm]{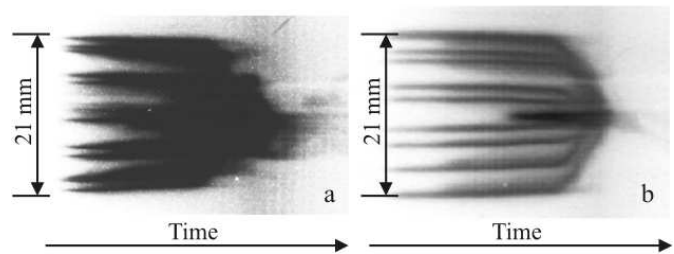}
  \end{center}
  \caption{Streak images: (a) an initially cold wire array implosion, (b) a wire temperature of 1.88~kK was maintained for one hour.}
  \label{wires}
\end{figure}

It is obvious from the streak images shown in Fig.~\ref{wires} that in the preheated wire array experiments, the wire array implosion dynamics is changed. As indicated earlier, in cold wire array experiments, the breakdown along a wire array occurs prior to melting of the wires. Further heating of unmelted wires is realized predominantly by heat conduction. Since this is a slow process, plasma formation is delayed and a radially expanded structure is produced that retards the implosion process. As it is shown in \cite{Pikuz:PRL:1999}, preheating of small-diameter wires has a pronounced effect on the fraction of the wire material passing into the plasma state and on the axial uniformity of the resulting plasma. This is sure to affect wire array implosion.

\section{Summary}

We have studied experimentally time variations of energy deposition and resistance of tungsten wires with 30.5 $\mu$m diameter for positive or negative polarity 
of the high-voltage electrode and varying initial wire temperature. Preheating is shown to increase the energy deposited in a wire during the resistive heating stage. 
The increase in energy deposition is due to an increase in the time interval between the onset of current flow and the electrode gap breakdown.

The energy deposited in a wire is found to depend on the polarity of the high-voltage electrode: for positive polarity the energy is $\sim 1.7$ times higher than for negative polarity. 

For negative polarity, breakdown is strongly connected with an emergence of 
emission centers on the cathode surface 
and with an uniform high-intensity light emission region on the wire surface $\sim 4$--6~mm far from the cathode.
For positive polarity, an emission center strongly connected with
the development of the breakdown appears on the wire surface $\sim 3$~mm far from the cathode.
	
The specific characteristics of the electrical explosion of fine wires in a wire array are identical to those of the explosion of a single wire. Preheating enables the wire array implosion dynamics to be changed and the fraction of high-energy radiation from a wire array to be increased.

\onecolumn

\clearpage

\section*{LIST OF FIGURES}

\end{document}